\documentclass[letter,comsoc]{IEEEtran}

\usepackage{cite}
\usepackage[T1]{fontenc}
\usepackage{graphicx}
\usepackage{subfigure}
\usepackage{graphics}
\usepackage{algorithm}
\usepackage{algorithmic}
\usepackage{amssymb}
\usepackage{amsthm}
\usepackage{color}
\usepackage{bm}

\usepackage[algo2e,linesnumbered,ruled]{algorithm2e}

\ifCLASSINFOpdf
\else
\fi
\usepackage{amsmath}

\usepackage[cmintegrals]{newtxmath}

\begin{document}
	
\title{Interference-Aware Deployment for Maximizing User Satisfaction in Multi-UAV Wireless Networks}

\author{Chuan-Chi~Lai,~\IEEEmembership{Member,~IEEE,}~Ang-Hsun~Tsai,~\IEEEmembership{Member,~IEEE,} Chia-Wei~Ting, Ko-Han~Lin, Jing-Chi~Ling, and~Chia-En~Tsai%
	\IEEEcompsocitemizethanks{
		\IEEEcompsocthanksitem This research was supported by National Science and Technology Council, Taiwan, R.O.C. under the Grant NSTC 110-2222-E-035-004-MY2. \emph{(Corresponding author: Chuan-Chi Lai)}
		\IEEEcompsocthanksitem Chaun-Chi Lai, Chia-Wei Ting, Ko-Han Lin, Jing-Chi Ling, and Chia-En Tsai are with the Department of Information Engineering and Computer Science, Feng Chia University, 40724 Taichung, Taiwan.
		\IEEEcompsocthanksitem Ang-Hsun Tsai is with the Department of Communications Engineering, Feng Chia University, 40724 Taichung, Taiwan.
		\IEEEcompsocthanksitem Corresponding author's e-mail: chuanclai@fcu.edu.tw.
	}
}

\markboth{Preprint for IEEE Wireless Communications Letters}
{Lai \MakeLowercase{\textit{et al.}}: Bare Demo of IEEEtran.cls for IEEE Communications Society Journals}


\maketitle

\begin{abstract}
In this letter, we study the deployment of Unmanned Aerial Vehicle mounted Base Stations (UAV-BSs) in multi-UAV cellular networks. 
We model the multi-UAV deployment problem as a user satisfaction maximization problem, that is, maximizing the proportion of served ground users (GUs) that meet a given minimum data rate requirement. 
We propose an interference-aware deployment (IAD) algorithm for serving arbitrarily distributed outdoor GUs. 
The proposed algorithm can alleviate the problem of overlapping coverage between adjacent UAV-BSs to minimize inter-cell interference. 
Therefore, reducing co-channel interference between UAV-BSs will improve user satisfaction and ensure that most GUs can achieve the minimum data rate requirement. 
Simulation results show that our proposed IAD outperforms comparative methods by more than 10\% in user satisfaction in high-density environments.
\end{abstract}


%
\IEEEpeerreviewmaketitle

\section{Introduction}
\label{intro}
\IEEEPARstart{U}{nmanned} Aerial Vehicle mounted Base Station (UAV-BS) can be used to build three-dimensional (3D) wireless network, and has become one of the important network carriers for 6G and Non-Terrestrial Networks (NTN)~\cite{9861699}~\cite{9456851}. UAV-BSs are not limited by the ground environment and can be used to dynamically deploy 3D wireless networks. 
Deploying UAV-BS at the appropriate altitude and horizontal position can provide better communication quality to terrestrial users through air-to-ground line-of-sight (LoS) propagation paths~\cite{6863654}. Therefore, UAV base stations are suitable for occasions that need to be deployed in a short period of time and dynamically deployed with the crowd to guarantee service quality.

Many existing works~\cite{8642333}~\cite{7762053}~\cite{9877894}~\cite{9177297} have proposed a variety of different UAV-BS deployment methods in wireless networks. In~\cite{8642333}, a density-aware deployment of single UAV-BS was proposed for serving arbitrarily distributed ground users (GUs). This method deploys single UAV-BS according to the density of GUs and ensures that the covered users meets the minimum data rate requirement.
In~\cite{7762053}, the proposed method deploys multiple UAV-BSs in a counterclockwise spiral from the map boundary to the center of the map until all GUs are covered. This approach is designed to minimize the number of UAV-BSs deployed, but lacks deployment flexibility as the altitude and service range of UAV-BSs are immutable. 
In~\cite{9877894}, the authors modeled the access control of GUs and UAV-BS deployment problem as a potential game and a convex optimization problem, respectively. 
They propose a method to examine a limited number of user association samples and then select the sample with the best response (or reward) as each ground user's access decision. Afterwards, based on the decisions of all users, they use a simple iterative algorithm to find the best position for the UAV-BS.
A data-driven placement (DDP) approach was proposed by~\cite{9177297}. DDP can decide the suitable number of UAV-BSs required, and then adjust the altitude and coverage of these UAV-BSs simultaneously, which can coexist with ground base stations (GBSs) to serve arbitrarily distributed GUs.

In contrast to the above existing works that focus on maximizing the system sum rate or coverage area, we focus on maximizing user satisfaction in a multi-UAV wireless network. User satisfaction is the proportion of GUs within a target area whose allocated data rate meets a given minimum requirement. The user satisfaction maximization problem considered is NP-hard and can be reduced from a \emph{0–1 Multiple Knapsack Problem}. 
In fact, we found that most unsatisfied users fell into coverage overlapping areas, which means that interference from adjacent UAV-BSs is a key challenge in the considered environment. Therefore, we propose a heuristic algorithm, the interference-aware deployment (IAD), to control the overlapping area between different UAV-BSs, thereby maximizing user satisfaction. In this letter, we consider an environment where the GUs arbitrarily distribute and density changes. The simulation results indicate that the proposed IAD outperforms the existing method in terms of user satisfaction.

The contributions of this work are summarized as follows:
\begin{itemize}
	\item We identify the user satisfaction maximization problem for multiple UAV-BS networks from the perspective of cellular operators. 
	\item We propose an Interference-Aware Deployment (IAD) approach that simplifies the considered user satisfaction maximization problem by replacing constraints related to Signal-to-Interference-plus-Noise (SINR) with tolerable distance control and adaptive association control.
	\item Unlike most existing works that consider relatively sparse GU distributions in typical uniform, Gaussian, or Poisson point process (PPP) distribution models, we focus on high-density scenarios with arbitrary and heterogeneous GU distributions. 
	\item The simulation results show that the proposed IAD outperforms the other existing methods in user satisfaction while GU density increases and the minimum data rate requirement increases. 
\end{itemize}

\section{System Model}
\label{systemmodel}
\begin{figure}[t]
	\centering
	\includegraphics[width=0.275\textwidth]{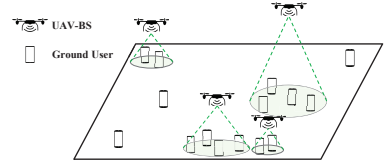}
	\caption{The considered system model.}
	\label{fig:SystemModel}
	\vspace{-5pt}
\end{figure}

As shown in Fig.~\ref{fig:SystemModel}, some outdoor activities are held in the serving area. The cellular operator uses $k$ homogeneous UAV-BSs, $\mathcal{U}=\{U_1,U_2,\dots,U_k\}$, to provide communication services, where $U_j=(u_j^x,u_j^y,u_j^h)$ is the 3D location of UAV-BS $U_j$ and $j=1,2,\dots,k$. We assume that $N$ GUs, $\mathcal{E}=\{E_1,E_2,\dots,E_N\}$, are arbitrarily distributed in the target area, where $E_i=(e_i^x,e_i^y)$ is the 2D location of ground user $E_i$ and $i=1,2,\dots,N$. Note that the user density is heterogeneous because different events are held at different locations.

For the channel model, a widely used air-to-ground pass loss model~\cite{6863654} is adopted. The probability of LoS signal from a UAV-BS $U_j$ to GU $E_i$ is as follows:
	\begin{align}\label{eq:PLos_PNLoS_to_user}
		P_{i,j}^\text{LoS}=\dfrac{1}{1+a\exp\left(-b(\theta_{i,j}-a)\right)},
	\end{align}
where $\theta_{i,j}=\frac{180}{\pi}\sin^{-1}(\frac{u_j^h}{d_{i,j}})$ is the elevation angle between GU $E_i$ and UAV-BS $U_j$; $u_j^h$ is the altitude of $U_j$; $a$ and $b$ are environmental constants related to the target area; and $d_{i,j}=\sqrt{(u_j^x-e_i^x)^2+(u_j^y-e_i^y)^2+(u_j^h)^2}$ is the Euclidean distance between $E_i$ and $U_j$. The probability of non-line-of-sight (NLoS) signals from UAV-BS $U_j$ to GU $E_i$ is $P_{i,j}^\text{NLoS}=1-P_{i,j}^\text{LoS}$.
The channel model of the connection between GU $E_i$ to UAV-BS $U_j$ with LoS and NLoS links is expressed as
\begin{align}\label{eq:model_los_nlos}
	L_{i,j}=&\begin{cases}
		20\log_{10}\left(\dfrac{4\pi f_c d_{i,j}}{c}\right)+\eta_{LoS}, & \text{LoS},\\
		20\log_{10}\left(\dfrac{4\pi f_c d_{i,j}}{c}\right)+\eta_{NLoS}, & \text{NLoS},
	\end{cases}
\end{align}
where $c$ is the speed of light; $f_c$ is the carrier frequency; and $\eta_{LoS}$ and $\eta_{NLoS}$ are the mean additional losses for LoS and NLoS. 
According to~\eqref{eq:PLos_PNLoS_to_user} and~\eqref{eq:model_los_nlos}, the average path loss between $E_i$ and UAV-BS $U_j$ can be expressed as
\begin{align}\label{eq:average_atg_model}
	\overline{L}_{i,j}&=P_{i,j}^\text{LoS}L_{i,j}^\text{LoS}+P_{i,j}^\text{NLoS}L_{i,j}^\text{NLoS}. 
\end{align}

We assume that all the UAV-BSs use a fixed transmission power $P^\text{T}$ to provide communication service. To successfully transmitting signals from UAV-BS $U_j$ to GU $E_i$, the signal-to-interference-plus-noise (SINR) of received signals at GU $E_i$ should be greater than a given threshold $\Gamma_{\rm th}$. Thus, the SINR value of $E_i$ associated with $U_j$ will be
\begin{align}\label{eq:sinr}
	\Gamma_{i,j}=\dfrac{P^\text{T}\cdot 10^{-\overline{L}_{i,j}/10}}{I_{\mathcal{U}\setminus \{U_j\}}+B_{i,j}N_0}\geq\Gamma_{\rm th}. 
\end{align}
where $I_{\mathcal{U}\setminus \{U_j\}}=\sum_{j'=1}^k P^\text{T}\cdot 10^{-\overline{L}_{i,j'}/10}\delta_{j,j'}$ is the interference power from the adjacent UAV-BSs if GU $E_i$ is in the coverage overlapping area, where $\delta_{j,j'}=1$ if $E_i$ locates in the overlapping coverage of UAV-BSs $U_j$ and $U_{j'}$, and $U_{j'}\in\mathcal{U},\forall j'\neq j$; otherwise, $\delta_{j,j'}=0$; $B_{i,j}$ is the allocated bandwidth (in Hz) of down-link connection from UAV-BS $U_j$ to a served GU $E_i$; $N_0$ is the power spectral density of the additive white Gaussian noise (AWGN). With the Shannon theorem and~\eqref{eq:sinr}, the allocated data rate (in bps) of GU $E_i$ associated with UAV-BS $U_j$ is
\begin{align}\label{eq:gu_data_rate}
	c_{i,j}=B_{i,j}\log_2\left(1+\Gamma_{i,j}\right).
\end{align}
According~\eqref{eq:gu_data_rate}, the sum rate of UAV-BS $U_j$ for serving its associated GUs is
\begin{align}\label{eq:uav_sum_rate}
	C_j=\sum_{i=1}^{N_j}c_{i,j}, 
\end{align}
where $N_j$ is the number of GUs associated with UAV-BS $U_j$.

\section{User Satisfaction Maximization Problem}
\label{sec:problem_formulation}
In this work, we consider the downlink performance of a multi-UAV wireless network. The considered system model is shown in Fig.~\ref{fig:SystemModel}. 
From the perspective of cellular operators, it is necessary to meet the minimum data rate requirement $c_{\min}$ of GU, which is predefined by cellular operators. Suppose $l$ levels of minimum data rate requirement, $C_{\rm level}=\{1*10^6,2*10^6,\dots,l*10^6\}$, are predefined and the cellular operator can select one level as the target minimum data rate requirement $c_{\min}$ to deploy UAV-BSs, where $c_{\min}\in C_{\rm level}$ and $l\in \mathbb{N^{+}}$. The deployment result must meet the needs of as many users as possible. In general, $c_{\min} \geq c_{\rm th}$ where $c_{\rm th}=B_{i,j}\log_2\left(1+\Gamma_{\rm th}\right)$  
	
With the aforementioned notation and assumptions, we call such a problem the \emph{user satisfaction maximization problem}, which can be expressed as
\begin{align}\label{eq:user_satisfaction_problem}
	\max_{\mathcal{U}}S= &\enspace \max_{\mathcal{U}}\dfrac{\sum_{j=1}^k\sum_{i=1}^{N}\psi_{i,j}}{N} \tag{P1}\\
	\text{subject to}&\enspace u_j^h\leq h_{\max}, \qquad\qquad\,\ \forall j,\label{eq:7}\\
	&\enspace r_j\leq r_{\max}, \qquad\qquad\;\,\ \forall j,\label{eq:8}\\
	&\enspace N_{\min}\leq N_j \leq N_{\max}, \quad \forall j,\label{eq:9}\\
	&\enspace \psi_{i,j}=\begin{cases}
		1 , & \text{if }c_{i,j}\geq c_{\min} \text{ and } r_{i,j}\leq r_j\\
		0, & \text{otherwise}	
	\end{cases}, \label{eq:10}\\
	&\enspace \sum_{i=1}^{N}\psi_{i,j}c_{i,j}\leq C_{\max}, \;\;\; \forall j,  \label{eq:11}\\
	&\enspace \sum_{j=1}^{k}\psi_{i,j}c_{i,j}\geq \sum_{j=1}^{k}\psi_{i,j}c_{\min}, \quad\,\forall i,\label{eq:12}\\
	&\enspace \sum_{j=1}^k\psi_{i,j}\leq 1,  \qquad\quad\;\;\; \forall i. \label{eq:13}
\end{align}
where $N$ is total number of GUs; $N_j=\sum_{i=1}^{N}\psi_{i,j}$ is the number of GUs associated with UAV-BS $U_j$; $r_{i,j}$ is the horizontal distance from GU $E_i$ to UAV-BS $U_j$; and $r_j$ is the coverage radius of UAV-BS $U_j$.

Constraint~\eqref{eq:7} guarantees that the deployed altitude $u_j^h$ of each UAV-BS does not excess the maximum altitude $h_{\max}$ which depends on the limitations of local laws and ability of UAV-BS. In the considered
system, we assume that the allowable path-loss of each GU is a fixed value, $L_{\rm allow}$. With $L_{\rm allow}$, the elevation angle $\theta_j^{\rm opt}$ can be obtained by solving the nonlinear partial differential equation $\frac{\partial r_{i,j}}{\partial\theta_j^{\rm opt}}=0$ of~\eqref{eq:average_atg_model}. Then, in constraint~\eqref{eq:8}, with the given $h_{\max}$, the maximum coverage radius of a UAV-BS $r_{\max}$ corresponding to $h_{\max}$ can be derived by $\theta_j^{\rm opt}=\tan^{-1}(\frac{h_{\max}}{r_{\max}})$.
Constraint~\eqref{eq:9} is used to ensure that the number of associations with UAV-BS is limited to a predefined range $[N_{\min},N_{\max}]$. In constraint~\eqref{eq:10}, $\psi_{i,j}$ is a binary function to indicate whether the allocated data rate of GU $E_i$ with respect to the associated UAV-BS $U_j$ can meet the minimum data rate requirement $c_{\min}$.
Constraint~\eqref{eq:11} is the admission control to ensure that the sum rate of served GUs does not exceed the backhaul capacity of the UAV-BS $C_{\max}$. For simplicity, $C_{\max}$ is set to a predefined value.
Constraint~\eqref{eq:12} is to guarantee that the minimum data rate requirement of a GU $E_i$ that is successfully associated with a UAV-BS $U_j$. Constraint~\eqref{eq:13} is used to ensure each GU is associated with at most one UAV-BS. Note that one GU may not be covered by any UAV-BS and thus $\sum_{j=1}^k\psi_{i,j}=0,\forall i$.

\section{The Proposed Interference-Aware Deployment}

\subsection{The Main Idea of IAD}
\subsubsection{Tolerable Distance Control}
\begin{figure}[!t]
	\centering  
	\includegraphics[width=0.285\textwidth]{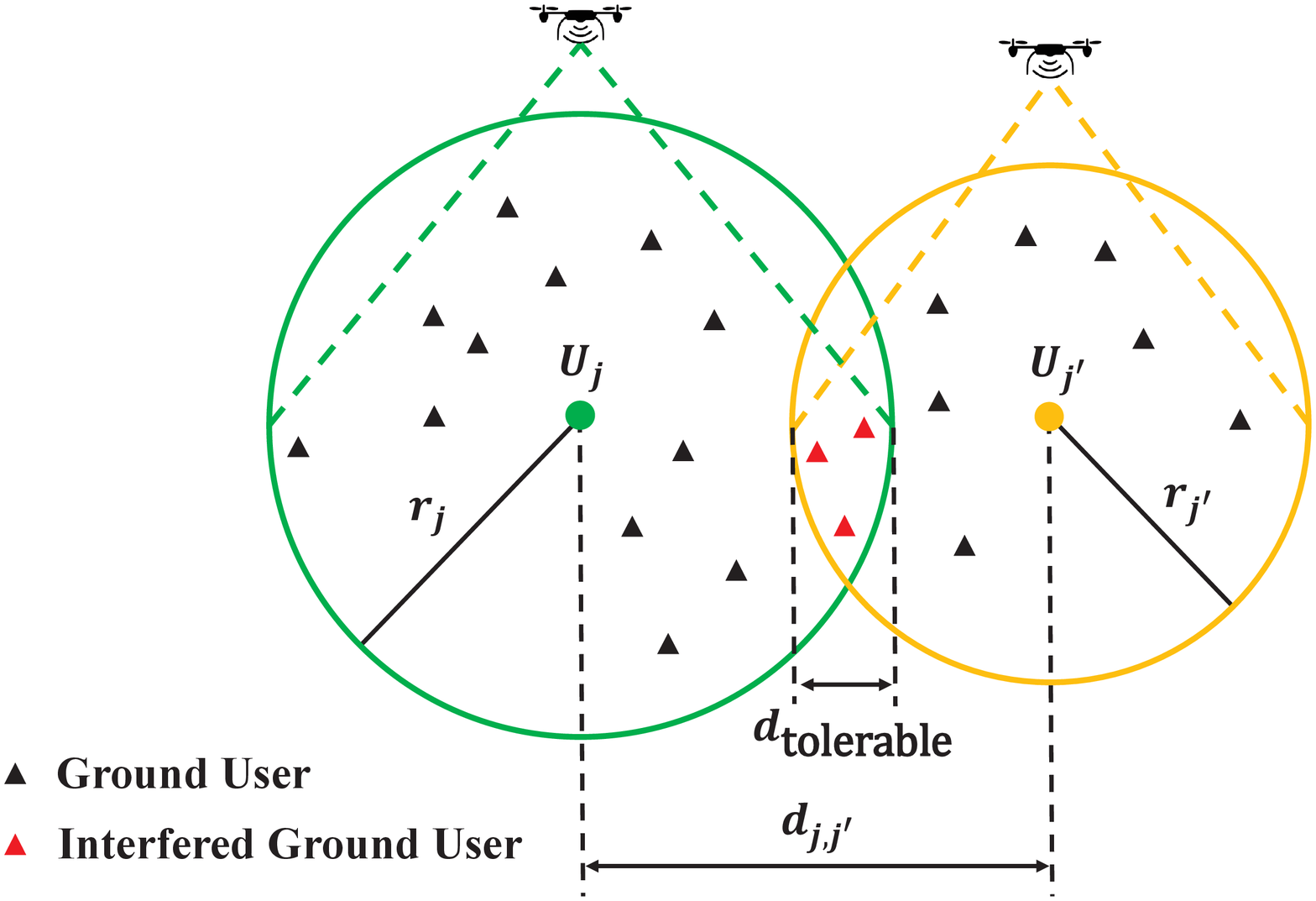}
	\caption{The design idea of IAD, where the tolerable distance $d_{\rm tolerable}$ is proposed to control the size of the overlapping area.}
	\label{fig:main_idea:d_tolerable}
	\vspace{-5pt}
\end{figure}

To maximize the user satisfaction of a multi-UAV wireless network, the proposed IAD aims to minimize the number of interfered GUs. Fig.~\ref{fig:main_idea:d_tolerable} shows the design idea of IAD, where black points are normal GUs and red points in the overlapping area are interfered GUs. Our proposal is to define a tolerable distance $d_{\rm tolerable}$ to control the size of the overlapping area. 
In IAD, the system will sequentially deploy UAV-BSs. If some UAV-BSs are already deployed in some 3D locations, IAD will prune a lot of search space to find the location of the next UAV-BS by using a filter condition until all the UAV-BSs are used up or no suitable location is found. Such a filter condition is defined as follows:
\begin{align}\label{eq:tolerable_distance}
	\left(d_{j,j'}>r_j+r_{j'}\right)||\left(r_j+r_{j'}-d_{j,j'}<d_{\rm tolerable}\right)\&\& \nonumber\\
	\left(d_{j,j'}> r_j\right)\&\&\left(d_{j,j'}> r_{j'}\right),
\end{align}
where $r_j$ and $r_{j'}$ are the coverage radii of two adjacent UAV-BSs $U_j$ and $U_{j'}$, $\forall j, j'=1,2,\dots,k, j\neq j'$, and $d_{j,j'}$ is the horizontal distance between two adjacent UAV-BSs.

In fact,~\eqref{eq:tolerable_distance} is designed to relax the constraints~\eqref{eq:10} and~\eqref{eq:12}, thus simplifying problem~\eqref{eq:user_satisfaction_problem}. Since satisfying the constraints~\eqref{eq:10} and~\eqref{eq:12} requires checking the interference received by each GU from all UAV-BSs to compute the corresponding SINR value, the computational complexity is relatively high. If we use~\eqref{eq:tolerable_distance} to search the candidates of UAV-BS deployment instead, the problem does not need to calculate the SINR value of each GU and the computational complexity can be significantly reduced.

\subsubsection{Adaptive Association Control}
In the considered problem~\eqref{eq:user_satisfaction_problem}, the target performance metric, user satisfaction, is related to the number of GUs having satisfied allocated data rates. However, as shown in constraint~\eqref{eq:11} each UAV-BS can only provide a limited total capacity $C_{\max}$, so the number of GU associations with each UAV-BS needs to be controlled. Otherwise, if there are too many GUs associated with the same UAV-BS, the allocated data rate of each GU may not be able to pass the minimum data rate requirement $c_{\min}$. Hence, in our proposed algorithm, we use the following condition to check whether the candidate deployment/association of each UAV-BS $U_j$ is valid or not. 
\begin{align}\label{eq:association_control}
	N_{\min}\leq N_{j}=\sum_{i=1}^N\phi_{i,j}\leq \left\lfloor \frac{C_{\max}}{c_{\min}} \right\rfloor =N_{\max}, \forall j,
\end{align}
and
\begin{align}\label{eq:16}
	\phi_{i,j}=\begin{cases}
		1 , & \text{if } r_{i,j}\leq r_j\\
		0, & \text{otherwise.}	
	\end{cases}
\end{align}
In other words, constraints~\eqref{eq:9} and~\eqref{eq:11} can be simplified to the condition~\eqref{eq:association_control} and the program can search the candidate position and associations of each UAV-BS without computing the exact SINR value of each GU.

\begin{algorithm2e}[!t]
	\footnotesize
	\SetAlgoVlined
	\SetKwComment{Comment}{/*}{*/}
	\caption{The main procedure of IAD, $\mathtt{IAD}()$}
	\label{alg:iad}
	\KwIn{$L_{\rm allow}, \mathcal{E}, C_{\max}, c_{\min}, h_{\max}, N_{\min}, d_{\rm tolerable}, k, \rho, m$}
	\KwOut{$\mathcal{U}, R$}
	Let $\mathcal{U}\leftarrow\emptyset$ be a global list to record deployed UAV-BSs $U_j=\{u_j^x,u_j^y,u_j^h\}, \forall j=1,2,\dots,k$\label{alg:iad:1}\;
	Let $R\leftarrow\emptyset$ be a global list to record the radii of UAV-BSs\label{alg:iad:2}\;
	Let $D_1,D_2\leftarrow\emptyset$ be three temporary candidate lists\label{alg:iad:3}\;	
	Randomly select a unlabeled GU $E_1$ from $\mathcal{E}$, find $E_1$'s two nearest neighboring unlabeled GUs $E_2$ and $E_3$ from $\mathcal{E}$, derive their circumcenter as the candidate location $l_{\rm cand}=(x,y)$, and then calculate the circumcircle as the candidate coverage radius $r$\label{alg:iad:4}\;
	\If{All of GUs in $\mathcal{E}$ are allocated/labelled or $|\mathcal{U}|$==k}{\label{alg:iad:5}
		\Return $\mathcal{U}, R$\label{alg:iad:6}\;
	}
	Let $N_{\max}\leftarrow\left\lfloor\frac{C_{\max}}{c_{\min}}\right\rfloor$\label{alg:iad:7}\;
		\tcp*[h]{invoke Algorithm~\ref{alg:allocation}}\;
		$[N^{\rm allocated}, R']\leftarrow \mathtt{Allocation}(\mathcal{E}, l_{\rm cand}, r_{\max}, N_{\max})$\label{alg:iad:8}\;
		\uIf{$(N^{\rm allocated}\geq N_{\min})\&\&(\mathcal{U}==\emptyset||$\eqref{eq:tolerable_distance} with respected to all candidate UAV-BSs in $\mathcal{U}$)}{\label{alg:iad:9}
		\For{$i=1;i \le m;i++$}{ \label{alg:iad:10}
			Find the fourth GU $E_4$ from the unassigned GUs in $\mathcal{E}$ that is the closest to $l_{\rm cand}$\label{alg:iad:11}\;
			Use the combination of three GUs choosing from $E_1,E_2,E_3,E_4$ to obtain four candidate coverage circles, and store them in $D_2$\label{alg:iad:12}\;
			\ForEach{element $e$ in $D_2$}{\label{alg:iad:13}
				\tcp*[h]{invoke Algorithm~\ref{alg:allocation}}\;
				$[N^{\rm allocated}, R']\leftarrow \mathtt{Allocation}(\mathcal{E}, e, r_{\max}, N_{\max})$\label{alg:iad:14}\;
				\If{$(N^{\rm allocated}\geq N_{\min})\&\&(\mathcal{U}==\emptyset||$\eqref{eq:tolerable_distance} with respected to all candidate UAV-BSs in $\mathcal{U}$)}{\label{alg:iad:15}
					save $e$ with $\mathtt{max}(R')$ into $D_1$\label{alg:iad:16}\;
					$l_{\rm cand}\leftarrow e$\label{alg:iad:17}\;
				}
			}
			$D_2\leftarrow\emptyset$\label{alg:iad:18}\;
		}
	}
	\Else{
		Jump to line~\ref{alg:iad:4}\label{alg:iad:20}\;
	}	
	\uIf{$|D_1|>0$}{\label{alg:iad:22}		
		$[l_{\rm cand},r_{\rm cand}]\leftarrow$ Choose the candidate that has the largest coverage radius from $D_1$\label{alg:iad:23}\;
		Add $l_{\rm cand}$ to $\mathcal{U}$ and add $r_{\rm cand}$ to $R$\label{alg:iad:24}\;
		Label the GUs associated with UAV-BS $l_{\rm cand}$ in $\mathcal{E}$ and then get the number of allocated GUs $N^{\rm allocated}$ in $\mathcal{E}$\label{alg:iad:25}\;
		\uIf{$|\mathcal{U}|<k$}{\label{alg:iad:26}
			\tcp*[h]{recursively invoke Algorithm~\ref{alg:iad}}\;
			$\mathtt{IAD}()$ with the same input parameters\label{alg:iad:27}\;
		}\Else{\label{alg:iad:28}
			\Return $\mathcal{U},R$\label{alg:iad:29}\;
		}
	}\Else{\label{alg:iad:30}
		Jump to line~\ref{alg:iad:4}\label{alg:iad:31}\;		
	}
\end{algorithm2e}

\subsection{Problem Transformation}
With the above two proposed ideas, the target problem~\eqref{eq:user_satisfaction_problem} can be transform to a simplified version of user satisfaction maximization problem which can be expressed as follows:
\begin{align}\label{eq:simplified_user_satisfaction_problem}
	\max_{\mathcal{U}}&\dfrac{\sum_{j=1}^k\sum_{i=1}^{N}\phi_{i,j}}{N} \tag{P2}\\
	\text{subject to}&\enspace \eqref{eq:7},~\eqref{eq:8},~\eqref{eq:tolerable_distance},~\eqref{eq:association_control},~\eqref{eq:16}\nonumber\\
	&\enspace \sum_{j=1}^k\phi_{i,j}\leq 1,  \qquad\quad\;\;\; \forall i. \label{eq:17}
\end{align}
With the help of the proposed association control and $N_{\min}$ and $c_{\min}$ are given in advance, the constraint~\eqref{eq:10} is also simplified as~\eqref{eq:16} without checking SINR values. Constraint~\eqref{eq:17} is corresponding to constraint~\eqref{eq:13}.

\subsection{The Procedure of IAD}
Algorithm~\ref{alg:iad} shows the main procedure of IAD and Algorithm~\ref{alg:allocation} shows the user allocation of a candidate UAV-BS. 
In Algorithm~\ref{alg:iad}, the required input information includes $L_{\rm allow}$, $\mathcal{E}$, $c_{\min}$, $C_{\max}$, $h_{\max}$, $N_{\min}$, $k$, $\rho$, and $m$, where $m$ is the limit on the number of iterations to find candidates. The outcomes of this algorithm are sets $\mathcal{U}$ and $R$.
From steps~\ref{alg:iad:1} to~\ref{alg:iad:3}, the IAD prepares some data structure to record importance information for searching the deployment decisions. Two temporary lists, $D_1$ and $D_2$, are used to help filter candidate deployments.
At step~\ref{alg:iad:4}, the IAD uses unlabeled GUs in $\mathcal{E}$ to find an initial candidate coverage as the deployment of a UAV-BS $l_{\rm cand}$.
Step~\ref{alg:iad:5} is one of the IAD's end point. If all GUs are already allocated or all UAV-BSs are exhausted, IAD will terminate at step~\ref{alg:iad:6}.
Step~\ref{alg:iad:7} is to determine the association constraint of~\eqref{eq:association_control} with the given $C_{\max}$ and $c_{\min}$.
At step~\ref{alg:iad:8}, the IAD calls Algorithm~\ref{alg:allocation} with $l_{\rm cand}$ to search the appropriate coverage radius, the covered GUs, and the number of associations of $l_{\rm cand}$. And then store the result into $D_1$ with $\mathtt{max}(R')$, where $R'$ stores the candidate radii and $\mathtt{max}(R')$ outputs the maximum coverage radius in $R'$. Step~\ref{alg:iad:9} is used to check whether the candidate can guarantee the constraints~\eqref{eq:tolerable_distance} and~\eqref{eq:association_control} of problem~\eqref{eq:simplified_user_satisfaction_problem}. If not, the procedure will back to step~\ref{alg:iad:4} to choose another one initial candidate.
From steps~\ref{alg:iad:10} to~\ref{alg:iad:18}, the IAD tries to find more possible candidates with the corresponding $\mathtt{max}(R')$ and store them into $D_1$. Within these steps, constraints~\eqref{eq:7} and~\eqref{eq:8}, are also guaranteed.
Step~\ref{alg:iad:22} checks if any candidates were found in this run. If no, the IAD will back to step~\ref{alg:iad:4}.
Steps~\ref{alg:iad:23} to~\ref{alg:iad:25} find the best candidate from $D_1$ and label all the allocated GUs. 
This condition of step~\ref{alg:iad:26} checks if all UAV-BSs are exhausted.
If there are any alternate UAV-BSs, the IAD will recursively call itself to continue searching for the next UAV-BS's candidate deployment at step~\ref{alg:iad:27}; otherwise, the IAD will terminate at step~\ref{alg:iad:29}. 
Since the IAD deploys UAV-BSs one-by-one recursively with uncovered GUs, the constraints~\eqref{eq:16} and~\eqref{eq:17} are also guaranteed.

\begin{algorithm2e}[!t]
	\footnotesize
	\SetAlgoVlined
	\caption{$\mathtt{Allocation}(\mathcal{E}, l_{\rm cand}, r_{\max}, N_{\max})$}
	\label{alg:allocation}
	\KwIn{$\mathcal{E}, l_{\rm cand}, r_{\max}, N_{\max}$}
	\KwOut{$N^{\rm allocated}, R'$}
	Let $R'$ be an ascending-order priority list based on the distance value\label{alg:allocation:1}\;
	$N^{\rm allocated}\leftarrow 0$\label{alg:allocation:2}\;
	\For{$i=1;i\le |\mathcal{E}|;i++$}{ \label{alg:allocation:3}     
		$R^{\rm temp}[i]\leftarrow \sqrt {{{(l_{\rm cand}.x - E_i.x)}^2} + {{(l_{\rm cand}.y - E_i.y)}^2}}$\label{alg:allocation:4}\tcp*[r]{$\forall E_i\in \mathcal{E}$}
		\If{$R^{\rm temp}[i]\leq r_{\max}$}{\label{alg:allocation:5}
			add $R^{\rm temp}[i]$ to $R'$\label{alg:allocation:6}\;
			$N^{\rm allocated}\leftarrow N^{\rm allocated}+1$\label{alg:allocation:7}\;
		}	
	}
	\If{$N^{\rm allocated}>N_{\max}$}{\label{alg:allocation:8}
		$R'$.remove($N_{\max}$,$|R'|-1$)\label{alg:allocation:9}\;
		\Return $N_{\max}, R'$\label{alg:allocation:10}\;
	}
	\Return $N^{\rm allocated}, R'$\label{alg:allocation:11}\;
\end{algorithm2e}


\begin{figure*}[!t]
	\centering	
	\subfigure[Varying $d_{\rm tolerable}$]{
		\label{fig:result:d_t} 
		\includegraphics[width=0.245\textwidth]{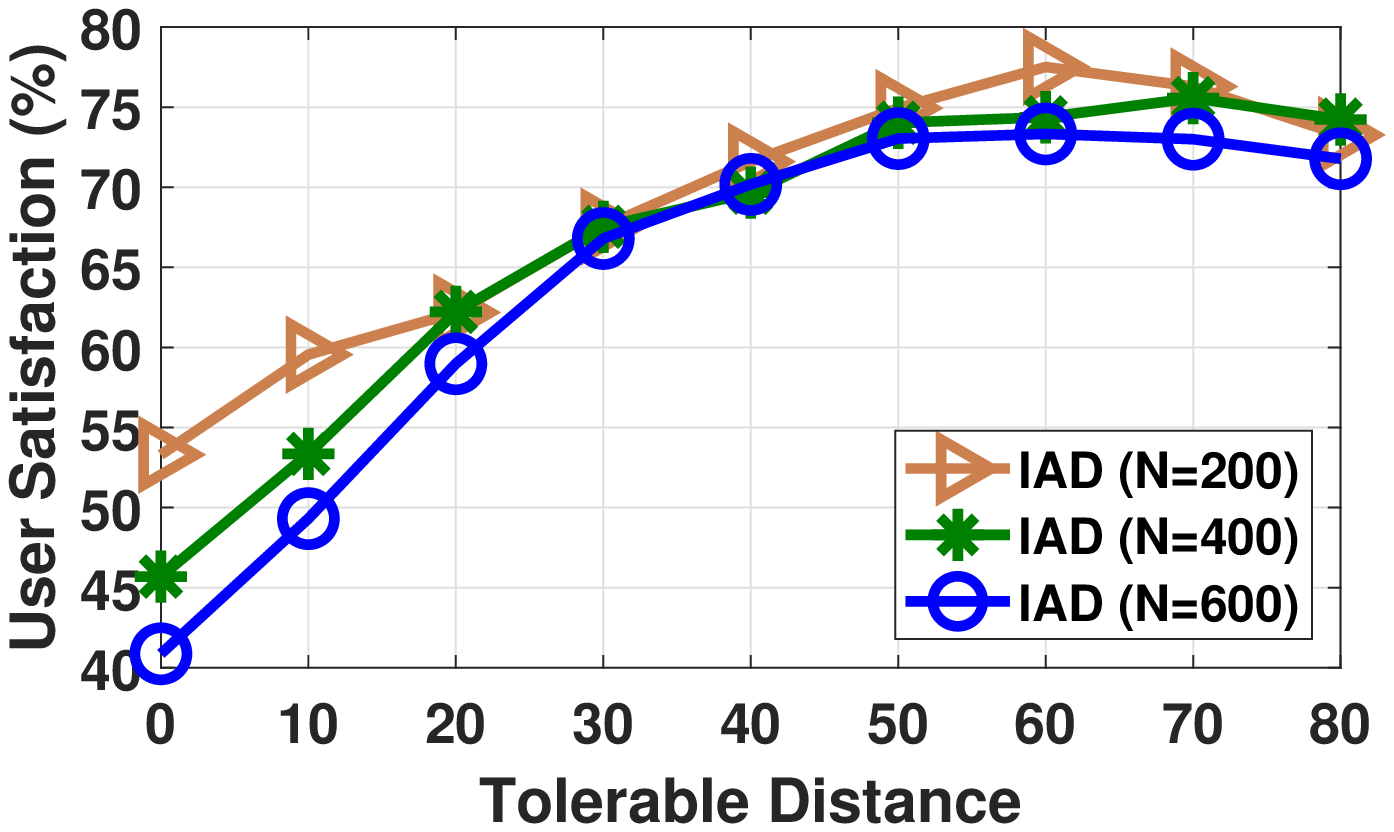}
	}\enspace	
	\subfigure[Varying $N$]{
		\label{fig:result:n_gu} 
		\includegraphics[width=0.245\textwidth]{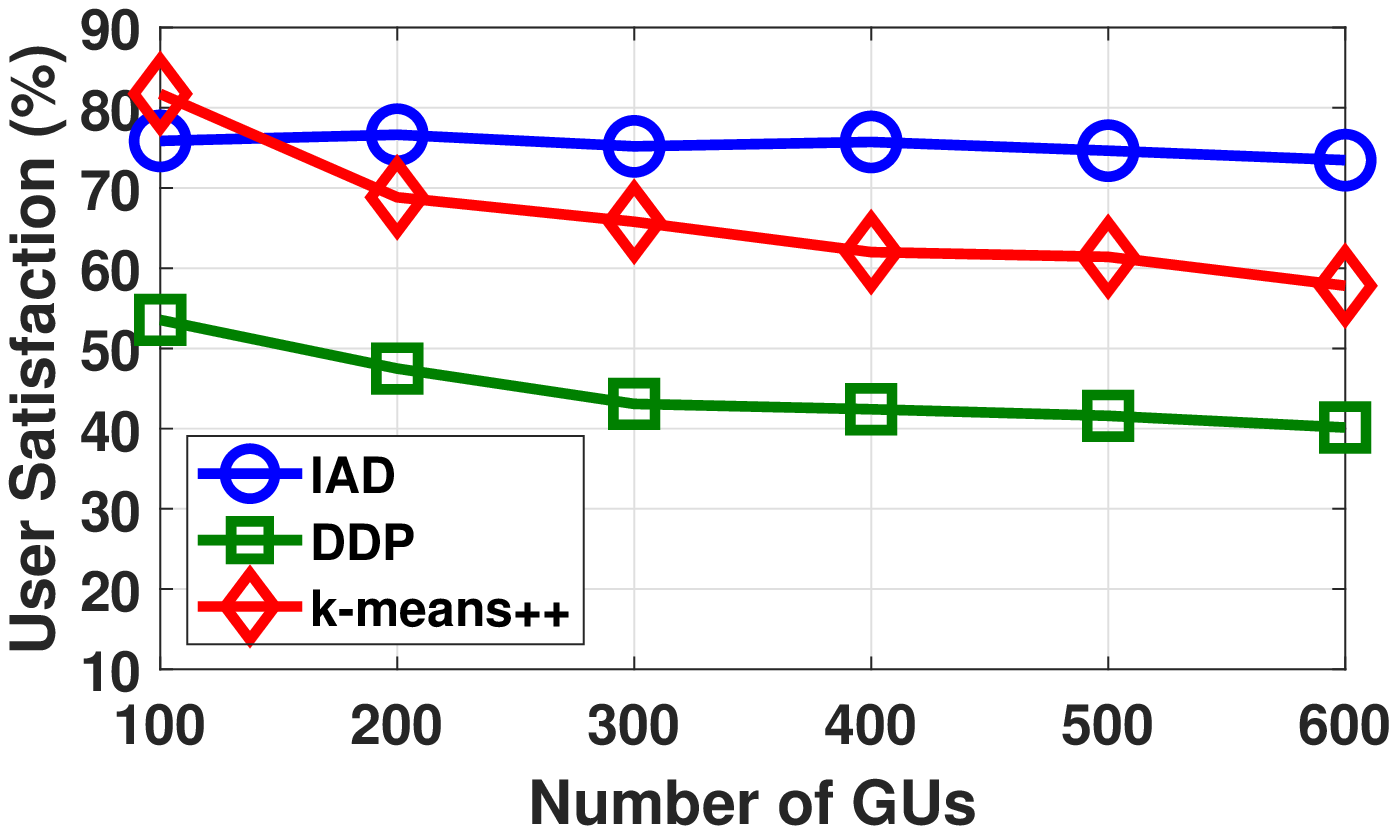}
	}\enspace
	\subfigure[Varying $c_{\min}$]{
		\label{fig:result:c_min} 
		\includegraphics[width=0.245\textwidth]{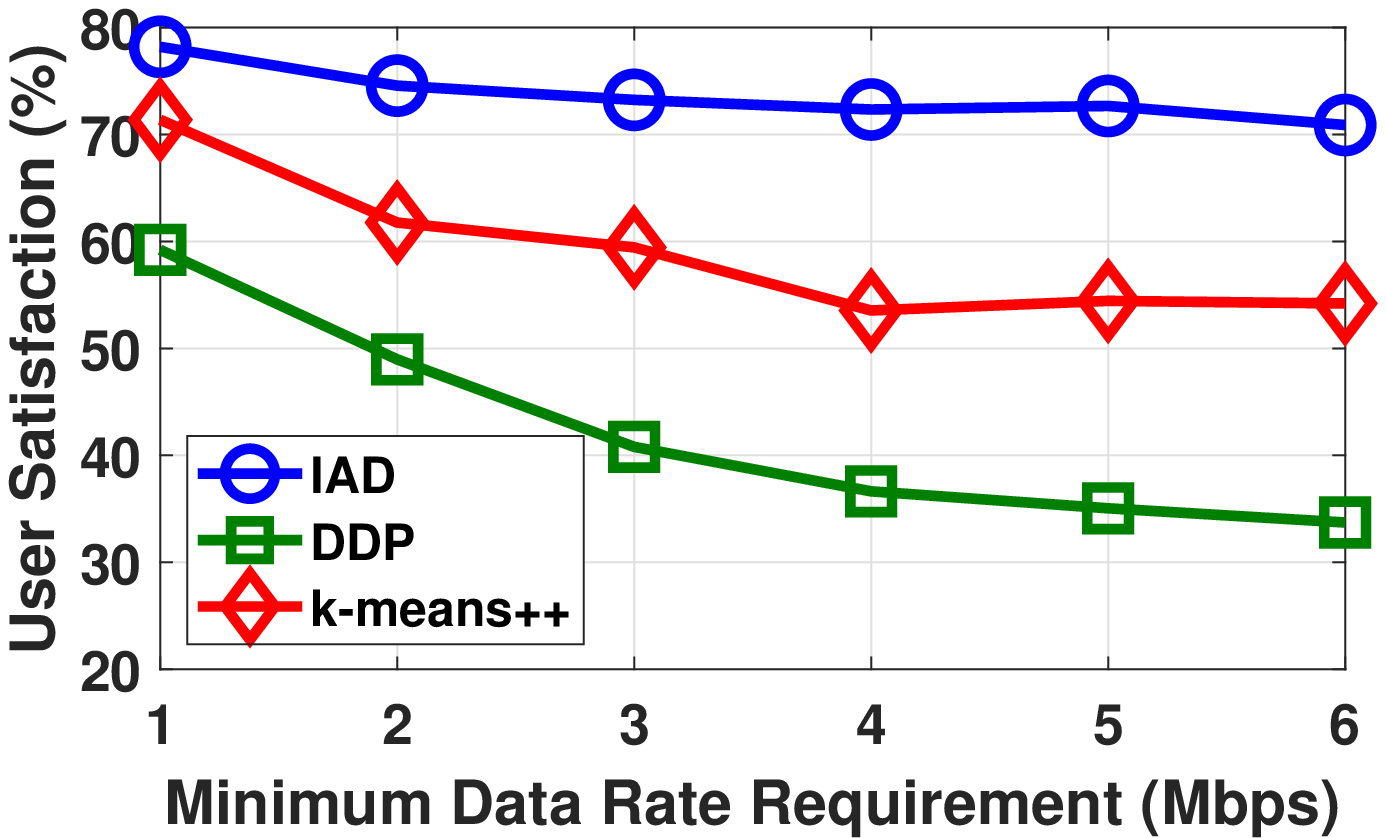}
	}%
	\caption{The performance results in terms of user satisfaction, $S$, while~\subref{fig:result:d_t} varying $d_{\rm tolerable}$ with $N=600$ and $c_{\min}=3$ Mbps,~\subref{fig:result:n_gu} varying $N$ with $d_{\rm tolerable}=60$ m and $c_{\min}=3$ Mbps, and~\subref{fig:result:c_min} varying $c_{\min}$ with $N=600$ and $d_{\rm tolerable}=60$ m.
	}
	\label{fig:result}   
	\vspace{-10pt}
\end{figure*}

\subsection{Complexity Comparison and Discussion}
The considered user satisfaction maximization problem~\eqref{eq:user_satisfaction_problem} is modeled as a knapsack-like problem with SINR-related constraints.
If we solve the problem~\eqref{eq:user_satisfaction_problem} directly, the simulation program needs to handle the SINR value of each GU and continuously update the interference from the other non-associated UAV-BSs while changing the candidate 3D locations of UAV-BSs. In general, such a straightforward way costs at least $O(kN^2)$.
One comparative method, DDP~\cite{9177297}, costs $O(kN^3)$ due to the use of Hungarian algorithm for balancing associations of UAV-BSs. Another method, SPIRAL~\cite{7762053}, aims the minimize the number of deployed UAV-BSs without considering any SINR-related constraints and takes $O(N^2\log N)$ time.
Compare to the aforementioned solutions, the proposed IAD algorithm can search the deployment of UAV-BSs without computing the SINR values of all GUs. That is, the proposed IAD can solve problem~\eqref{eq:simplified_user_satisfaction_problem} in $O(kN)$ time.

\section{Simulation Results and Discussions}
\label{simulation}
In this section, we discuss the system performance in terms of user satisfaction, $S$, while varying the tolerable distance, $d_{\rm tolerable}$, number of GUs, $N$, and the minimum data rate requirement, $c_{\min}$. The simulations are implemented by MATLAB R2022a. We assume that there are all the GUs arbitrarily distributed in a $600\times 600$ m$^2$ dense urban area and the densities of GUs are heterogeneous. 
All performance results are mean output values simulated under 100 different GU distributions.
To validate the performance of IAD, we conduct the following methods for comparison: $k$-means++, DDP~\cite{9177297} and IAD (our method).
	
The parameters of the air-to-ground channel model are $(a, b, \eta_{\rm LoS}, \eta_{\rm NLoS})=(12.08, 0.11, 1.6, 23)$ as adopted for the dense urban environment in~\cite{7510820}. According to the local laws~\cite{laws_taiwan}, $h_{\max}$ is set to 120 meters. The other remaining parameters of our simulation are set by default to $f_c=2.4$~GHz, $c=3\times 10^8$~m/s, $B=2\times10^7$~Hz, $N_0=-174$~dBm/Hz, $P^{\rm T}=20$~dBm, $\Gamma_{\rm th}=5$~dB, $L_{\rm allow}=119$~dB, $C_j=1.5\times10^8$~bps, $N_{\min}=10$, and $k=25$. With~\eqref{eq:PLos_PNLoS_to_user} to~\eqref{eq:average_atg_model} and the above given parameters, the optimal elevation angle of each UAV-BS is $\theta_j^{\rm opt}=54.69^{\circ}$ and the maximum coverage of each UAV-BS is $r_{\max}=h_{\max}\tan^{-1}\theta_j^{\rm opt}=85$ m.
	
We first observe the effect of varying $d_{\rm tolerable}$ on user satisfaction performance under the cases of $N\in\{200,400,600\}$ and $c_{\min}=3$ Mbps. As shown in Fig.~\ref{fig:result:d_t}, the performance trends of different cases are similar and increase as $d_{\rm tolerable}$ becomes large. The IAD seems to have convergent performance results near $80\%$ user satisfaction around $d_{\rm tolerable}=60$~m. This result is the mean of input samples from 100 different GU distributions. In fact, the optimal $d_{\rm tolerable}$ is highly related to the GU distribution. Unlike the typical uniform, Gaussian, or Poisson point process (PPP) distribution models discussed in most existing works, we use samples from arbitrary GU distributions as input in this letter, so it is difficult to formulate $d_{\rm tolerable}$ in a closed-form.
	
Second, we vary the number of GUs, $N$, from 100 to 800 to observe user satisfaction performance of all comparative methods under the case of $d_{\rm tolerable}=60$ and $c_{\min}=3$ Mbps. Fig.~\ref{fig:result:n_gu} shows that the proposed IAD outperforms $k$-means++ and DDP~\cite{9177297} by more than 10\% and 30\% respectively when $N$ becomes large. On the contrary, $k$-means++ can outperform IAD only in a sparse GU environment. If the distribution of GUs become dense, both DDP and $k$-means++ will have serious inter-cell interference. It shows that the design of tolerable distance control~\eqref{eq:tolerable_distance} can effectively reduce the overlapping coverage. As a result, more GUs experience less interference and are allocated a good enough data rate, which can then effectively improve user satisfaction. Compared with DDP and $k$-means++, IAD is a more suitable solution for dense GU scenarios, such as outdoor concerts or New Year's Eve parties.
	
Finally, we set $l=6$ to vary $c_{\min}$ from $1$ Mbps to $6$ Mbps to observe the user satisfaction performance under the case of $N=600$ and $d_{\rm tolerable}=60$. Fig.~\ref{fig:result:c_min} shows that the proposed IAD outperforms the other methods in all cases of $c_{\min}$. As $c_{\min}$ increases, the user satisfaction of IAD only decreases slightly and this is benefited from the proposed adaptive association control. Therefore, compared with DDP and $k$-means++, IAD can provide the best UAV-BS deployment for high data rate.

\section{Conclusion}
\label{conclusion}
In this letter, we investigated how to deploy multiple UAV-BSs with controllable interference to serve arbitrarily distributed users. We modelled the considered multi-UAV deployment problem as a user satisfaction maximization problem and proposed two non-SINR related constraints for problem simplification. Then, we proposed the interference-aware deployment (IAD) algorithm to solve this simplified problem. According to simulation results, IAD can effectively alleviate the overlapping coverage problem between adjacent UAV-BSs to minimize inter-cell interference, and maintain a reasonable association on each UAV-BS to ensure the minimum data rate requirement of most GUs, so as to maximize user satisfaction. In particular, the IAD outperforms the existing methods by more than 10\% when the density of GUs becomes crowded.


\bibliographystyle{IEEEtran}
\bibliography{IEEEabrv,reference}

\end{document}